\documentclass[conference]{IEEEtran}

\IEEEoverridecommandlockouts
\usepackage{xcolor}
\usepackage{comment}
\usepackage[utf8]{inputenc}
\usepackage{graphics}%, graphicx}
\usepackage{setspace}
\usepackage{amsmath,amsfonts}
\usepackage{times}
\usepackage{color}
\usepackage{subfigure}
\usepackage{amssymb}
\usepackage{eurosym}
\usepackage{textpos}
\usepackage[normalem]{ulem}
\usepackage{multirow}
\usepackage{wrapfig}
\usepackage[dvipdfmx]{graphicx}
\usepackage{mathtools}
\usepackage{siunitx}
\usepackage{algorithm2e}

\DeclareMathOperator*{\minimize}{minimize}

\def\logg{\mathrm{log}}

\def\argmax{\operatornamewithlimits{arg\,max}}

\def\GHz{~\mathrm{GHz}}
\def\MHz{~\mathrm{MHz}}
\def\Hz{~\mathrm{Hz}}
\def\Mbps{~\mathrm{Mbps}}
\def\m{~\mathrm{m}}

\def\dB{~\mathrm{dB}}
\def\dBm{~\mathrm{dBm}}

\newcommand{\mathrmbold}[1]{\boldsymbol{\mathrm{#1}}}
\def\params{\mathrmbold{w}}

\def\tablescale{0.915}

\begin{document}
\RestyleAlgo{ruled}

%\title{Deployment of % Placement of Multiple 
%Aerial Mobile Access Points and User Associations in Highly Dynamic Environments
%}
\title{Deployment of Aerial Mobile Access Points and User Association in Highly Dynamic Environments}

\title{QoS-aware User Association, Number, and 3D Placement of Aerial Vehicles in dynamic Mobile Networks}

\title{Dynamic User Association and QoS-aware Number and 3D Placement of Aerial Vehicles in 6G Networks}

\title{QoS-aware Number and 3D Placement of 6G Aerial Vehicles via Dynamic User Association}

\title{Cost-efficient and QoS-aware User Association, Number and 3D Placement of 6G Aerial Vehicles}

\title{Cost-Efficient and QoS-Aware User Association and 3D Placement of 6G Aerial Mobile Access Points}

%\title{QoS-aware optimization of the number and 3D placement of aerial vehicles in dynamic UAV-assisted Network via cost minimization and user association}
%--------------------------
%  AUTHORS
%--------------------------

\author{\IEEEauthorblockN{Esteban Catté, Mohamed Sana and Mickael Maman}

\IEEEauthorblockA{ CEA-Leti, Universite Grenoble Alpes, F-38000 Grenoble, France \\
%CEA, LETI, Minatec Campus, 17 rue des Martyrs, 38054 Grenoble, France\\
\{esteban.catte, mohamed.sana, mickael.maman\}@cea.fr\\}
}

% make the title area
\maketitle

%--------------------------
%  ABSTRACT
%--------------------------
\maketitle

\begin{abstract}
 %trade off between resource efficiency, latency at the constraint of target Reliability are studied. The realization  of the proposed approach can be demonstrated by HARQ protocol.
%Mobile access points (MAP) are being studied for a wide spectrum of opportunities. On very dense cities with high resource demand, on uncovered area or along huge highways, MAPs can be deployed to provide fast assistance to the network. As the fifth generation of telecommunication (5G) become more heterogeneous, MAPs deployment and management become a major challenge. The key is to determine efficiently and quickly the position and the configuration of MAPs on a dynamic environment. To this end, we formulate the joint 3D MAP deployment and user association problem over a dynamic network under interference constraint. We propose an iterative probabilistic approach to place MAPs and a Deep Multi-Agent Reinforcement learning algorithm for the user-association. Our simulation results shows that our approach...

%The sixth generation (6G) of wireless 
6G networks require a flexible infrastructure to dynamically provide ubiquitous network coverage. %and efficiently extend the communication. 
Mobile Access Points (MAP) deployment is a promising solution. In this paper, we formulate the joint 3D MAP deployment and user association problem over a dynamic network under interference and mobility constraints. First, we propose an iterative algorithm to optimize the deployment of MAPs. Our solution efficiently and quickly determines the number, position and configuration of MAPs for highly dynamic scenarios. MAPs provide appropriate Quality of Service (QoS) connectivity to mobile ground user in mm-wave or sub-6GHz bands and find their optimal positions in a 3D grid. Each MAP also implies an energy cost (e.g. for travel) to be minimized. Once all MAPs deployed, a deep multi-agent reinforcement learning algorithm is proposed to associate multiple users to multiple MAPs %access points 
under interference constraint. Each user acts as an independent agent that operates in a fully distributed architecture and maximizes the network sum-rate.

\end{abstract}

\IEEEpeerreviewmaketitle

\section{Introduction}
The sixth-generation (6G) of wireless networks aims to dynamically and efficiently extend the communication environment to enable access to all people, information, and goods, anywhere and anytime, in an ultra-real-time experience. This requires the design and development of mechanisms for the dynamic coverage and connectivity extension through the exploitation of innovative devices (e.g., drones, robots, cars). 
These innovative devices can act in the network as Mobile Access Points to cover areas that are difficult to access, where the infrastructure is only needed for a limited and short time, or where the regular network infrastructure has been damaged. This flexible infrastructure \cite{Zeng2019} brings several challenges: dynamic three-dimensional (3D) MAPs deployment and their trajectory adaptation, line-of-sight management, and dynamical network management and configuration (e.g., associating multiple users with multiple access points).
%The fifth generation of mobile telecommunication is developed to provide connectivity for a wide range of applications. With more heterogeneity comes the development of mobile access points and their optimization. On very dense cities, on uncovered area or along huge highways, MAPs can be deployed to provide fast assistance to the network and must have self-configuration, self-healing and self-optimization mechanism \cite{Wang2019}. This new flexibility adds a three dimensional dynamic configuration, a new path adaptation constraint, line of sights problems and dynamical network management and configuration, i.e, with user association \cite{Zeng2019}.

The problem of MAP placement attracts particular attention in the literature. It is a challenging problem, involving complex path-planning, as well as radio resource optimization and network management.
%The placement of MAPs become notoriously difficult as the network management includes larger parameter set, radio access management, path-planning and complex optimization criteria. 
%Optimally solving such a problem generally gives arise to the following problems i) how many MAPs to deploy, and ii) how and where to deploy them with respect to the network dynamic.
The optimal solution of such a problem usually gives rise to the following problems: i) how many MAPs to deploy, and ii) how and where to deploy them with respect to the network dynamics.
In this context, a genetic algorithm is proposed in \cite{Lim2021} for network coverage extension. In \cite{Peer2020}, the authors designed a framework for drone placement under user mobility constraints. Authors in \cite{Sharadeddine2018} adopts a two-phase approach, iteratively optimizing the number of Unmanned Aerial Vehicles (UAVs) to deploy and adjusting their positions with respect to (w.r.t.) user equipment's(UE) locations. A similar approach is proposed in \cite{sabzehali2021} based on graph theory taking into account backhaul constraint. However, none of these solutions jointly consider the dynamic of user traffic request, their mobility, and their association with MAPs during optimization, which is the focus of our paper. 
To address this problem, authors in \cite{ghanavi2018efficient} proposed a reinforcement learning algorithm that continuously learns and adapts the placement of a MAP w.r.t. users mobility in order to maximize the network sum-rate. In \cite{Li2021}, the authors jointly optimize the coverage extension, together with UAV trajectory and spectrum allocation, but without mobility. However, all of these approaches are applied to a single UAV and have failed to generalize to multiple MAPs, which is a difficult problem. Specifically, a key task in MAP placement optimization is to dynamically determine the optimal UE association with multiple MAPs. Inefficient user association can severely affect network spectral efficiency and UEs' perceived QoS. Unfortunately this problem cannot be reduced to connecting users to the nearest Base Station (BS) due to co-channel interference and unfavorable channels conditions, which impact network performance. %The state-of-the-art does not really take into account user associations constraints while the position of the UAV can be strongly impacted by these constraints. 
To address this issue, a machine learning algorithm and a contract theory algorithm are proposed in \cite{2Zhang2021} to solve the MAP placement problem while guaranteeing user QoS. However, this work does not consider mobility. %Users and each deployed UAV make a contract to offload traffic by guarantying a level of performance in hot spot area but without considering mobility. 
In \cite{Li2019}, the authors proposed dual connectivity management where users can be connected with the ground BS and MAPs simultaneously using a genetic algorithm to optimize network sum-rate. In \cite{Sun2020}, the authors proposed a joint optimization of 3D MAP placement and user association but for a single UAV. Similarly, \cite{kalantari2017user} found the MAP position, user association, and backhaul configuration using a clustering algorithm. However, these solutions do not consider user mobility, highly impacting the interference network profile. In our previous work \cite{Sana2020}, we proposed a Deep Multi-Agent Reinforcement Learning (MARL) based user association algorithm, which jointly considers co-channel interference and network traffic dynamics. Moreover the solution can handle a variable number of users and their mobility. In solution proposed in this work, each user acts as an independent agent operating in a fully distributed architecture and making autonomous decisions. %Based on this approach, 
We propose a two-phase approach, which first optimizes the number and placement of MAPs to minimize the associated deployment cost. We achieve this first step with a \emph{scalable iterative Monte Carlo} based algorithm, which jointly optimizes the number and position of deployed MAPs. Next, in the second step, we optimize the user association to provide access to UEs in both sub-6GHz and and mm-wave band and to guarantee the UEs' QoS requirement. 
Our proposed solution jointly considers inter- and intra-cell interference, user mobility, and user traffic request during the optimization. As a result, the proposed solution adapts well by design to dynamic networks with mobility, dynamic traffic, and varying number of UEs. 

The remainder of the paper is organized as follows. Section II presents the system model and formulates the addressed problem. Then, Section III describes the proposed solution, whereas Section IV provides numerical results, demonstrating the performance of our proposed algorithm. Finally, Section V concludes the paper.

\section{System Model and Problem Formulation}

\subsection{System Model}\label{sysmodel}

\begin{figure}[!t]
%\vspace{-0.3cm}
\centering
\includegraphics[width=\columnwidth]{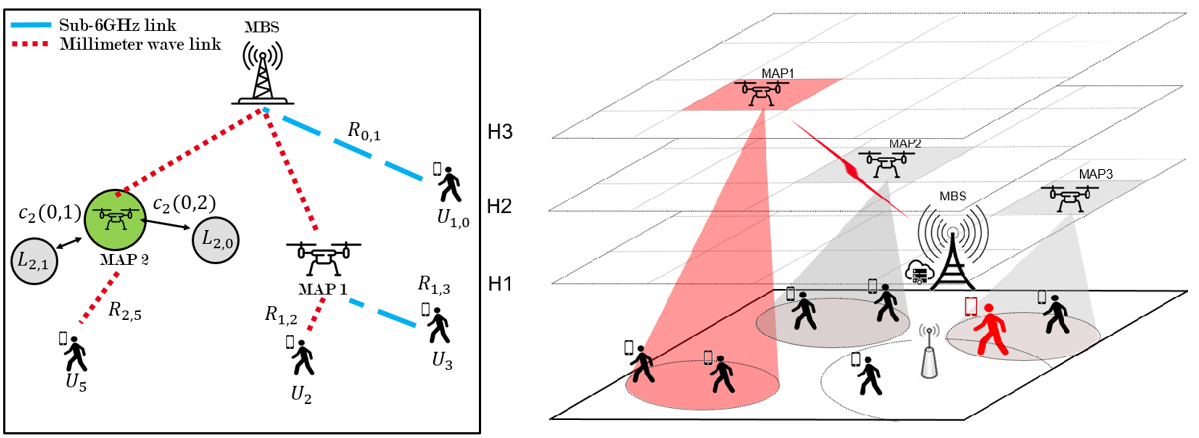}
\vspace{-0.6cm}
\caption{Cell architecture and 3D cell configuration}
%\vspace{0.05cm}
\vspace{-0.3cm}
\label{fig:cellscheme}
\end{figure}

We consider a downlink network where $K$ MAPs, (e.g., UAVs), are jointly deployed with a Macro Base Station (MBS) to provide ubiquitous coverage to $P$ UEs. We define $\mathcal{A}$ the set of APs and $\mathcal{U}= \{u_0,u_1,...,u_{P-1}\}$ the set of UEs. 
We assume that each UE $i\in\mathcal{U}$ is equipped with two antennas and can communicate at sub-6GHz and mm-wave frequencies with the MBS and MAPs, respectively. % to a subset $\mathcal{U}_{i}$ of UEs. 
In this network, we focus on the optimization of MAP locations jointly with Radio Resource Management (RRM). In this context, let $l_{i,k} = \{x_{i,k},y_{i,k},z_{i,k}\}$ denote $k$-th possible location of the MAP $i$ represented as a coordinate in 3D dimensional space as represented in Fig. \ref{fig:cellscheme}. 
%\sout{Let $\phi_{i}$ be the deployment strategy for MAP $i$. We denote with $\mathcal{L}(\phi_{i})=\{l_{i,k}\}_{k=0,\dots,L_i-1}$ the set of all $L_i$ possible locations of MAP $i$ given its strategy $\phi_{i}$.}
We denote with $\mathcal{L}_i=\{l_{i,k}\}_{k=0,\dots,L_i-1}$ the set of all possible locations of MAP $i$, and let $\mathcal{L}=\cup_{i\in\mathcal{A}}\mathcal{L}_i$ denote the set of all possible MAPs locations, which can be determined \emph{a priori} using path-planning or defined as the set of possible safe locations of MAPs e.g., in urban area. We define $\ell_{i,k}$ as the binary variable which equals $1$ if the MAP $i$ is effectively deployed on its $k$-th location and $0$ otherwise. Moving a MAP from one location to another incurs a certain cost either in terms of energy consumption, network operation or renting. We consider this aspect by defining $c_i(k,p)$  as the cost associated to moving a MAP $i$ from location $k\in\mathcal{L}_i(t)$ to location $p\in\mathcal{L}_i(t+1)$:
\begin{equation}\label{costi}
    c_i(k,p) = e_i(k,p) E_c+ c_{i,0},
\end{equation}
where $E_c$ is the cost of a unit of energy, $e_i(k,p)$ is the energy consumed by MAP $i$ to move from $k$ to $p$, which is a function of the distance \cite{Abeywickrama2018}, and $c_{i,0}$ is a fixed cost due to, e.g., the renting of MAP $i$. 
%\sout{The energy consumption depends on the type of MAP: connected cars or UAVs do not imply the same energy model. \textcolor{red}{Il faut ne parler que de UAV oublie les cars ...}}
Hence, we can define the total cost $C(t)$ incurred by the deployment of all MAPs as follows:
\begin{equation}\label{eq:deploy_cost}
    C(t) = \sum _{i\in \mathcal{A}} \underbrace{\sum_{k\in \mathcal{L}_i(t)} \sum_{p \in \mathcal{L}_i(t+1)} c_i(k, p) \ell_{i,k}(t)\ell_{i,p}(t+1)}_{C_i(t)}.
\end{equation}
Such a cost function may also vary depending on the number of targeted UEs, which will be served by the MAP. Therefore, our first objective focuses on determining the optimal subset $\mathcal{D}\subset \mathcal{L}$ of MAPs locations that minimize $C(t)$ by jointly minimizing the number of deployed MAPs and optimizing their deployment \emph{w.r.t.} UEs' QoS. % of MAPs, and optimally associating UEs.
Accordingly, our second objective focuses on the user association problem. This is because the optimal assignment of UEs to APs improves the network spectral efficiency and the perceived QoS of UEs \cite{Sana2020}. Hence, let us denote with $x_{i,j}(t)$ the binary association variable, which equals $1$ if UE $j$ is associated with AP $i\in\mathcal{A}$ at time $t$, and $0$ otherwise. % (where $\mathcal{A}_{j}$ denotes the set of MAPs that UE $j$ can connect to). We assume that when UE cannot associate with any AP in $\mathcal{A}_j$, it  When $\mathcal{A}_j = \emptyset$, UE $j$ is connected to the MBS %and when $\mathcal{A}_j = {i} , \forall i \in \mathcal{A}$, the UE is connected to AP $i$ (if AP $i$ accepts the UE, otherwise the MBS). 
%The association problem comes when $\card(\mathcal{A}_j) > 1$. 
We assume that all APs perform beam training in advance, so that they are able to set up an appropriate beam when a connection is established between AP $i$ and UE $j$. We denote with $R_{i,j}(t)$ the corresponding communication rate, which is given by the Shannon capacity:
\begin{equation}\label{rate}
      R_{i,j}(t) = B_{i,j} \logg_{2}(1+\mathrm{SINR}_{i,j}(t)),
\end{equation}
where $B_{i,j}$ is the bandwidth allocated by AP $i$ to UE $j$ and $\mathrm{SINR}_{i,j}$ the signal-to-interference-plus-noise ratio between AP $i$ and UE $j$, which comprises intra-cell and inter-cell interference of both grounded and mobile APs.
Then, given the data demand of UE $j$, $D_j(t)$, we define its QoS's satisfaction $\kappa_j(t)\in[0,1]$ as follows:
\begin{equation}
    \kappa_{j}(t) = \sum_{i\in\mathcal{A}} x_{i,j}(t) \mathrm{min}\left(1, \frac{R_{i,j}(t)}{D_{j}(t)}\right).
\end{equation} 
Accordingly, we say that the QoS is fully satisfied when $\kappa_{j}(t)=1$. Finally, to account with fairness in the association, we define the total network utility function as:
\begin{equation}\label{sumrate}
    R_{\alpha}(t) = \sum_{i \in \mathcal{A}} \sum_{j \in \mathcal{U}} x_{i,j}U_{\alpha}\left(\min\left(D_j(t),R_{i,j}(t)\right)\right),
\end{equation}
%Knowing every user rate, we define the user quality of service $\kappa_{j}(t)\in [0,1]$ w.r.t its associated AP $i$ as:
%Moreover, we consider UE $j$ demand satisfaction $\kappa_{j}(t)\in [0,1]$ w.r.t. its associated AP $i$ as follows:
%The QoS is fully satisfied when $\kappa_{j}(t)=1$.
%Finally, with all UE connections, we define $R(t)$ as the total utility function of the network:
% \begin{align}\label{eq:utility}
%     R(t) = \sum_{i\in\mathcal{A}}\sum_{j\in\mathcal{U}} U_{\alpha}(R_{i,j}(t)),
% \end{align}
where $U_{\alpha}(\cdot)$ is the $\alpha$-fair utility function given in \cite{Altman2008} as:
\begin{equation}
	U_{\alpha}(x)={}\left \{
        \begin{array}{l l}
    	(1-\alpha)^{-1}{x^{1-\alpha}}, & \forall \alpha \geq 0 \text{~and~} \alpha \neq 1,\\
		\mathrm{log}(x), &\quad \text{if~} \alpha = 1.
						\end{array}
					\right.
\end{equation}

%================Cell scheme=========

\subsection{Channel Model}
The channel model varies according to several factors such as the radio environment (i.e. suburban, urban, dense urban, high rise building), the communication band (i.e. sub-6GHz and mm-wave), and the type of communication (i.e. ground-to-ground or ground-to-air). %, and the direction of the communication (i.e. uplink or downlink).
In general, the channel path loss ${\rm PL}_T$ for any communication link can be defined on the basis of the Line of Sight (LoS) conditions as follows:
\begin{equation}\label{pathloss1}
    {\rm PL}_T = p{\rm PL}_{Los} + (1-p){\rm PL}_{NLos},
\end{equation}
where $p$ is the LoS probability, ${\rm PL}_{Los}$, and ${\rm PL}_{NLos}$ are the  LoS and NLoS path loss respectively.\\[0.2em]
\textbf{Air/ground sub-6GHz path-loss.} Following \cite{Al-Hourani2014}, we define the LoS probability as a function of the elevation angle $\theta$ : 
\begin{equation}\label{plosairground}
    p(\theta) = c(\theta- \theta_{0})^d,
\end{equation}
where $\theta_0 = 15^{\circ}$ is the lowest possible angle and $c,d$ are environmental parameters, and we compute the frequency-dependent path loss model as a function of the link type $l\in\{\rm Los, NLos\}$:

\begin{equation}\label{pathloss2.1}
    {\rm PL}_{l} = 20\logg_{10}(d)+20\logg_{10}(f)-27.55 + \chi_{\sigma_l}.
\end{equation}
Here, $d$ the distance between the transmitter and the receiver, $f$ is the carrier frequency in $\mathrm{MHz}$, and $\chi_{\sigma_l}$ is the shadowing coefficient, which follows a normal distribution with a mean $\mu_{l}$ and a standard deviation $\sigma_{l}$, whose values are given in \cite{Al-Hourani2014}.\\[0.2em]
%and $\sigma_{NLos}$ where $\sigma_{Los/NLos}=a_{Los/NLos}e^{-b_{Los/NLos} \theta}$ with $a,b$ representing environmental parameters. The values of the model parameters for the different environments are provided by Al-Hourani et al. in \cite{Al-Hourani2014}.\\[0.2em]
\textbf{Air/ground mm-wave path-loss.} Here, we define the LoS probability as a function of the height of the transmitter ($h_t$) and receiver ($h_r$) and some environmental parameters \cite{Yi2019}:
\begin{equation}\label{plosairground4}
    p(d) = \prod _{n=0}^{\mathclap{\max(0,\gamma(d))}}1-e^{\left(-\frac{\gamma(d)\max(h_t,h_r)-(n+\frac{1}{2})(\left | h_t-h_r \right |)^2}{2\epsilon^2\gamma(d)^2}\right)}.
\end{equation}
In \eqref{plosairground4}, $\gamma(d)$ represents the average number of buildings crossing the link between the transmitter and the receiver separated by a distance $d$. 
%Also $\epsilon$ represents the random height of each obstacle, which follows a Rayleigh distribution. 
%\textcolor{red}{formule a checker. esque $\eta$, $\gamma$ represente le shadowing? }.
Hence, the distance-dependent path loss model is \cite{shakhatreh2021modeling}:
\begin{equation}\label{pathloss3}
    {\rm PL}_{l} = \alpha_{l} + 10\beta_{l}\logg_{10}(d)+\chi_{\sigma_l}; ~l\in\{\rm Los, NLos\},
\end{equation}
where, %$\gamma_{l}$ follows a normal distribution with a mean $\mu_{l}$ and a standard deviation $\sigma_{l}$, and 
$\alpha_l$, $\beta_l$ depend on the radio environment.\\[0.2em]
\textbf{Ground/ground sub-6GHz or mm-wave path-loss.} Here, the path loss model can be defined without considering the LoS probability \cite{Sun2016}:
\begin{equation}\label{pathloss_ground_ground}
    {\rm PL}_{T}(d) = 10\alpha \logg_{10}(d) + \beta + 10\gamma \logg_{10}(f) + \chi_{\sigma}.
\end{equation}
Where $d$ is the distance between the transmitter and the receiver, $f$ the carrier frequency and
$\chi_{\sigma}$ the shadowing effect.

\subsection{Formulation of MAP Deployment Problem}

%In that part, we define our MAP deployment problem for the presented system model. 
% Inside the network, MAPs must provide a connection to mobile ground UEs. Each MAP is moving within the cell according to predefined locations and moving from one location to another incurs a cost. We can define $C(t)$, the total cost to deploy a given set of MAPs as follows:
% %the instantaneous MAP deployment cost as follows:
% \begin{equation}
%     C(t) = \sum _{i\in \mathcal{M}} \underbrace{\sum_{k\in \mathcal{L}_i(t)} \sum_{p \in \mathcal{L}_i(t)} c_i(k, p) \ell_{i,p}(t-1) \ell_{i,k}(t)}_{C_i(t)}
% \end{equation}

% %For the given system model and problem formulation, the value $C(t)$ represent the total cost to deploy a given set of MAP. 
% It depends on the deployment cost of each MAP $i$,  $c_{i}(k,p)$, to move from position $k$ to position $p$. 
% We define the instantaneous cost for MAP $i$ as : 

% \begin{equation}\label{costi}
%     c_i(k,p) = (Ke_i(k,p)E_c+s),
% \end{equation}
% where $E_c$ is the cost of a unit of energy, $Ke_i(k,p)$ is the energy consumed by MAP $i$ to move from $k$ to $p$ and $s$ is a fixed cost (e.g., MAP rent or other external costs depending on the scenario).
%The energy consumption depends on the type of MAP: connected cars or UAVs do not imply the same energy model.
%\textcolor{red}{Il faut ne parler que de UAV oublie les cars ...}
After the above definitions, we formulate the MAP deployment problem to minimize the total deployment cost as:
\begin{align}
    \minimize_{x_{i,j}, \ell_{i,k}}~& \frac{1}{T}\sum_{t=0}^{T-1} C(t) \tag{$\mathcal{P}_1$} \label{eq:P1}\\
     \text{s.t.}~~ & x_{i,j}(t),~\ell_{i,k}(t)\in \{0,1\}, & \forall i,j,k,t
    \tag{$\mathcal{C}_1$} \label{eq:C1}\\
    &C_i(t) \leq C_{\max}, &\forall i\in\mathcal{A}\backslash\{0\}
    \tag{$\mathcal{C}_2$} \label{eq:C2}\\
    &\sum _{j\in \mathcal{U}} x_{i,j}(t) \leq N_{i}, & \forall i,t, \tag{$\mathcal{C}_3$} \label{eq:C3}\\
    &\sum _{i\in \mathcal{A}} x_{i,j}(t) = 1, ~~ & \forall j, t, \tag{$\mathcal{C}_4$} \label{eq:C4}\\
    &\kappa_j(t) \geq \mathcal{Q}_j, & \forall j,t, \tag{$\mathcal{C}_5$} \label{eq:C5}\\
    %&\textcolor{red}{P_{i,j}(t) \leq P_{\max},} & \forall i, j, t, \tag{$\mathcal{C}_{\rm 1f}$} \label{eq:C1f}\\
    &\sum_{\mathclap{k\in \mathcal{L}_i(t)}} \ell_{i,k}(t) \leq 1, & \forall t, i \in \mathcal{A}\backslash\{0\}, \tag{$\mathcal{C}_6$} \label{eq:C6}\\
    &\sum_{\mathclap{i\in\mathcal{A}\backslash\{0\}}}\quad\sum_{~~\mathclap{k\in \mathcal{L}_i(t)}} \ell_{i,k}(t) \leq K_{\max}, & \forall t, \tag{$\mathcal{C}_7$} \label{eq:C7}
    %&\mathcal{L}_i(t) \subset \mathcal{L}, & \forall t, i \in \mathcal{A}\backslash\{0\}. \tag{$\mathcal{C}_8$} \label{eq:C8}
\end{align}
The constraint \eqref{eq:C1} defines $x_{i,j}$ and $\ell_{i,k}$ as binary variables. The constraint \eqref{eq:C2} ensures that the deployment cost of a MAP is lower than the maximum cost $C_{\max}$. The constraints \eqref{eq:C3} and \eqref{eq:C4} ensure that each AP $i$ serves at most $N_i$ UEs and that each UE is associated to exactly one AP. The constraint \eqref{eq:C5} guarantees the QoS satisfaction of each UE. Finally, the constraints \eqref{eq:C6}-\eqref{eq:C7} guarantees that a MAP is deployed to at most one location at a time and that the total number of deployed MAP does not exceed $K_{\max}$. % at time $t$. The constraint \eqref{eq:C8} ensures that the MAP $i$ moves only to defined positions of $\mathcal{L}$.
% Note: we could consider instead of \eqref{eq:P1} the long-term cost: \[\lim_{T\to\infty}\frac{1}{T}\sum_{t=0}^{T-1} \mathbb{E}[C(t)]\]
%As a result, 
It is worth noting that Problem \eqref{eq:P1} is non-convex and NP-hard, thus difficult to solve with classical optimizations techniques.

\section{Proposed Solution}
\begin{figure}[!t]
%\vspace{-0.3cm}
\centering
\includegraphics[width=\columnwidth]{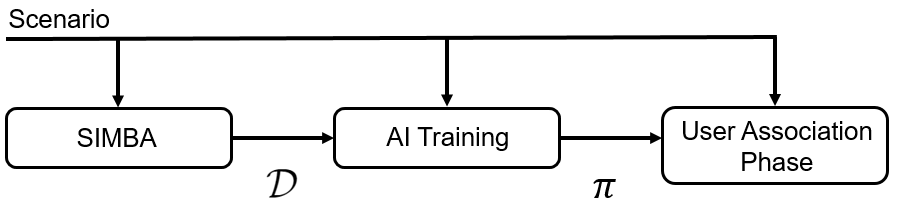}
\vspace{-0.6cm}
\caption{Proposed solution architecture}
\vspace{-0.3cm}
\label{fig:cellschemearchi}
\end{figure}

%ce qu'on propose
Our proposed solution for deploying MAPs jointly considers the deployment cost, UEs' mobility, co-channel interference, and traffic request dynamic. %, which strongly affect the system's performance. %In this study, MAPs deployment considers UE mobility, deployment cost, interference and different channel types.
One key challenge is that the optimal MAP deployment strategy strongly depends on UEs' traffic requests and the co-channel interference that will be generated, which is not known until UEs are fully associated. At the same time, the optimal association of UEs also depends on the MAP deployment. This ping-pong effect makes the problem very complex and difficult to solve. %Thus we hinge on %The UE channel model includes neighbourhood interference, which means that UEs must first be associated to learn the interference profile of the network and then associate to the best MAP.
%Since deployment depends on UE demand, MAPs need to be aware of interference within the network. Our approach considers and solves both feedback loops iteratively.
%Réduire un peu la complexité avec positions prédéfinies => On se place dans un cas discret
%petite figure
To limit such a complexity, we first define a 3D grid of positions for MAPs. The discretization of the 3D space gives a finite number of solutions for the problem \ref{eq:P1}. However, the search space of possible solutions remains large, prohibiting any exhaustive search approach. 
%the number of possible location and MAP number leads to a gigantic number of combination, prohibiting exhaustive research. 
Thus, we design \texttt{SIMBA}, a \emph{Scalable Iterative Monte-Carlo Based Algorithm}, with low-complexity, which explores the search space to find (sub)-optimal solutions as illustrated in in Fig. \ref{fig:cellschemearchi}. \texttt{SIMBA}, first performs Monte-Carlo explorations of MAPs deployment strategies and then exploits the best solution by adopting  %deploys scenario-aware MAPs and uses 
a standard user association algorithm based on maximum Signal-to-noise-ratio (MAX-SNR) to find the sub-optimal MAP deployment with low complexity. Next, based on \texttt{SIMBA} output, we apply our previously proposed MARL framework to train a user association, which in contrast to the MAX-SNR algorithm, considers co-channel interference. We show that this approach is able to compensate the sub-optimality of \texttt{SIMBA}. %and reduce imperfections due to spatial discretization.   % policy  network performs UE association training with knowledge of the AP locations. The AI algorithm reduces imperfections due to spatial discretization and corrects ambiguous association situations of MAX-SNR approach that does not consider interference.

%================Solution architecture=========
%==========================================================

\subsection{\texttt{SIMBA}: Scalable Iterative Monte-Carlo Based Algorithm}
%Map deployement
This section describes \texttt{SIMBA}, a low-complexity iterative algorithm (see Algorithm \ref{alg:simba}), which finds near-optimal MAPs deployment solution (in terms of deployment cost and UEs' QoS satisfaction). %This method aims at a near-optimal deployment, in terms of deployment cost, with low complexity (Algorithm \ref{alg:simba}).
\texttt{SIMBA} alternates a Monte-Carlo exploration and exploitation phases. During an exploration, \texttt{SIMBA} randomly samples a set $\mathcal{M}_k$ of $k$ locations on which MAPs are deployed. Each time it deploys a MAP $i$ on a location $p$, it updates a score associated to this location. Let $\mathrm{score}_p^{(i)}(t)$ be such a score:
%\sout{the algorithm randomly deploys each MAP on one of its potential locations $\mathcal{L}_{i}$ including its charging station. The number of successfully deployed MAPs gives the number of drones and the set of chosen locations gives their position in the cell. Let $\mathrm{score}_i(t)$ be the score associated to the location where drone $i$ is positioned:}
%\begin{equation}\label{score}
    %Sc_{L_{i}}(t) =  \frac{1}{U'} %\min(1,\frac{R_{i,j(t)}}{D_{j}(t)})
%\end{equation}
\begin{equation}\label{score}
    \mathrm{score}_p^{(i)}(t) = \frac{1}{|\mathcal{U}_i|}\sum_{j\in\mathcal{U}_i} \kappa_j(t),
\end{equation}
where $\mathcal{U}_i$ is the number of UEs served by MAP $i$.
%Pourquoi le score ? Comment il est calculé avant
As we consider interference and mobility, the score of a position is computed based on UEs' perceived QoS. Thus, higher the score, better the QoS of UEs served by a MAP deployed at that position. % and not just the number of associated UEs to take advantage of MAPs on a dense 6G network.
In the exploitation phase, a MAP is deployed at the location with the highest score. This location is no longer sampled until a solution of \ref{eq:P1} has been found. Then we iterate over $T$ episodes at the end of which we select the best solution of MAPs deployment $\mathcal{D}$. % When the set of fixed MAPs solves the problem \ref{eq:P1}, the best solution is updated. 
Eventually, given $\mathcal{D}$, we learn the optimal user association using a multi-agent reinforcement learning based approach. %we perform After defining the positions and the number of MAPs to deploy, AI agents are trained to correct the interference and space discretization imperfections. 
%\begin{algorithm}
%\caption{SIMBA Deployment algorithm}\label{alg:simba}
%\SetKwFunction{Fct}{Deploy}
%\KwIn{$\mathcal{L}$ $\leftarrow$ Set of positions}
%$U \leftarrow$ Set of users \\ 
%$D \leftarrow$ Number of draws \\
%\KwOut{$A \leftarrow$ Number of MAPs and selected positions} 
%\SetKwProg{Fn}{Function}{:}{}
%\Fn{\Fct{}}{
%Init MAP set\\
%\While {$t<T$}{
%\For{$d \in D$}{
%\For{$i \in \mathcal{M}$}{
%    Draw next position
%  }
%Update users radio resources and position\\
%User association procedure\\
%\For {$l \in \mathcal{L} $}{
%Compute scores}
%\EndFor
%}
%Fix a MAP to the highest score position\\
%Add the MAP to the set of deployed drones\\
%Compute the cost $C(t)$ and $\kappa_j(t), ~\forall j$\\
%\If{\eqref{eq:P1} is solved}{
%\If{cost < best solution cost}{
%Save current deployment
%Reset MAP set\\}}
%$t=t+1$
%}
%\Return The number of MAPs, their positions and deployment cost\\
%}
%\end{algorithm}
\begin{algorithm}[!h]
%\SetKwFunction{Fct}{Deploy}
\caption{\texttt{SIMBA} MAP Deployment Algorithm}\label{alg:simba}
\KwIn{
Define the set of MAPs possible locations $\mathcal{L}$ and the number of Monte-Carlo iterations $M$.
}
Initialize the score of locations: $\mathrm{score}_p=0, ~\forall p\in\mathcal{L}$.\\
Initialize the set of deployed MAP locations $\mathcal{D}=\emptyset$.\\ 
%Define the score vector $S=\{s_p\}_{p=1\dots|\mathcal{L}|}$ of size $|\mathcal{L}|$.\\
%\SetKwProg{Fn}{Function}{:}{}
%\Fn{\Fct{}}{
Set $C_{\min} = \infty$ and $k=K_{\max}$ (max. number of UAVs)\\
%Set \\
\For {$t=1, \dots, T$}{
Initialize an empty set of locations $\mathcal{D}_t=\emptyset$.\\
\For{$s=1\dots K_{\max}$} {
\emph{\textbf{Step 1:} Monte-Carlo exploration}\\
\For{$m=1,\dots,M$}{
    Randomly sample $k$ locations $\mathcal{M}_k(m) \sim \mathcal{L}$.\\
    Deploy a UAV to location $p,~\forall p \in \mathcal{M}_k(m)$.\\
    Perform user association procedure$^1$.\\
    \For {$p \in \mathcal{M}_k(m) $}{
        Compute the score of $p$ using Eq. \eqref{score}.\\
        Update $\mathrm{score}_p$ of location $p$.
    }
}
\emph{\textbf{Step 2:} Monte-Carlo exploitation}\\
Store location $i=\argmax_i\{\mathrm{score}_i\}$ in $\mathcal{D}_t$.\\
Deploy UAVs into the locations in $\mathcal{D}_t$.\\
%Fix a MAP to the location with highest score\\
%Add the MAP to the set of deployed drones\\
Compute $C(t)$, and $\kappa_j(t), ~\forall j$.\\
\eIf{\eqref{eq:C1}-\eqref{eq:C7} are guaranteed}{
Break.\\
}{
Remove $i$ from $\mathcal{L}$ and set $k=k-1$.\\
%$C_{\min} = C(t)$
%Remove $i$ from $\mathcal{D}$.
}
}
\emph{\textbf{Step 3:} Test Monte-Carlo solution}\\
\If{$C(t) < C_{\min}$}{
$C_{\min} = C(t)$\\
Save current deployment $\mathcal{D}=\mathcal{D}_t$
}
%Update UEs mobility and channels\\
%$t=t+1$\\
}
%\Return The number of MAPs, their positions and deployment cost\\
%}
\KwOut{$\mathcal{D}$ set of locations of deployed MAPs} 
{\footnotesize $^1$\textbf{Note}: Here, we adopt the MAX-SNR algorithm to limit complexity.}
\end{algorithm}
\vspace{-0.3cm}

\subsection{User Association}\label{user_association}
In this section, we describe the proposed MARL algorithm for user association. In the proposed framework, we model each UE as an agent, which cooperatively learns with its teammates a common user association policy through interaction with the shared radio environment. To this end, agents learn to map their local and global observations $o_j(t)$ of the radio environment to actions $a_j(t)$ corresponding to connection requests towards MAPs. Following our previous work \cite{Sana2021}, let $o_{j}^{l}(t)=\{R_{\alpha}(t), {\rm RSS}_j(t), {\rm AoA}_j(t),R_{a_j(t),j}(t),D_j(t)\}$ denote the local observation, which comprises the received signal strength $\mathrm{RSS}_j$, and the associated angle of arrival $\mathrm{AoA}$ w.r.t to all MAPs. Here, $R_{a_j(t),j}(t)$ represents UE $j$'s perceived rate and $D_j(t)$ the UE $j$ data demand. Moreover let $o_{j}^{g}(t)= \{(x_k(t), y_k(t), z_k(t), a_k(t-1)), ~\forall k\in\mathcal{N}_j(t)\}$ denote the global observations of UE $j$, where $(x_k(t), y_k(t), z_k(t))$ is the location of its $k$-th neighbors, $a_k(t-1)$ is the connection request in previous time slot, and $\mathcal{N}_j(t)$ denotes the UE neighborhood. This global observations represent UE $j$ perception of its surrounding environment. The goal of the learning procedure is to define the user association policy $\pi_{\params}$, with learnable weights $\params$, which outputs the association probability vector $p_j(t) = \pi(o_{j}(t))\in \mathbb{R}^{|\mathcal{A}|}$ that maximizes the sum of $\gamma$-discounted rewards over a time horizon $T_e$:
\begin{equation}\label{discount_reward}
    G_j(t) = \sum _{\tau=t+1}^{T_{e}} \gamma^{\tau-t-1}R_{\alpha}(\tau),
\end{equation}
where $\gamma$ is the discount factor such that $0<\gamma<1$. Finally, we construct the policy $\pi_{\params}$ using an actor-critic module, which is optimized via \emph{proximal policy optimization} \cite{Sana2021}. In particular, our proposed solution is specifically conceived to handle dynamic networks with varying number and position of UEs.

\subsection{Complexity Analysis}
As we discretize the 3D-space, a naive algorithm may find the optimal solution of problem \ref{eq:P1} using an exhaustive search. This algorithm has a complexity of $\mathcal{O}(C)$, where $C=\sum _{i=0}^{K_{\max}}\begin{pmatrix} |\mathcal{L}|\\ i \end{pmatrix}$ is the number of possible combinations of locations. In the worse case scenario where $K_{\max} \geq \frac{|\mathcal{L}|}{2}$, we have $C\geq 2^{|\mathcal{L}|-1}$. Note that each combination is not guaranteed to solve \ref{eq:P1}, which also requires solving the user association, leading to a prohibitive solution with very high complexity. In contrast, \texttt{SIMBA} has a complexity of approximately $\mathcal{O}(T M K_{\max})$, which scales linearly with $T$ and the number of Monte-Carlo iterations $M$ that we can conveniently choose to find (sub)-optimal solutions in a reasonable time.
%In contrast, the complexity of our algorithm is related to the number of MAPs, UEs and the MAX-SNR association algorithm. %and is strongly influenced by the interference computation. 
%In our simulations, interference is calculated for each UE using a random set of neighbors to ensure constant complexity.\footnote{The closest UEs are not considered because a far connection can cause a lot of interference with other mobile devices and not all UEs are considered to reduce computational complexity}.

%The complexity of our MAX-SNR algorithm depends on the UEs' SNR quick sort algorithm, used to chose the AP with the best channel configuration. We approximate this complexity by $C_{MAX-SNR}=P\logg(P)$.

%Let $M$ be the number of Monte-Carlo iterations of \texttt{SIMBA}. In the worse case scenario, considering a complexity of $O(M(P\logg{P}+4K))$ for the exploration phase, $O(P\logg{P})$ for the exploitation phase and $O(4)$ for the solution test, the complexity of our algorithm is $C=O(((P\logg{P}+4K)M+P\logg{P})K_{\max}+4)T))$.

%Let $\nu$ be the number of neighbors considered to compute interference and $M$ the number of Monte-Carlo iterations of our algorithm. Considering a complexity of $O(M(P\logg{P}+\nu\logg{\nu}+3K))$ for exploration phase, $O(\nu\logg{\nu}+P\logg{P}+4)$ for the exploitation phase and $O(K)$ for each iteration, the complexity of our algorithm is $C=O(((M(P\logg{P}+3K)+\nu\logg{\nu}+P\logg{P}+4)K+3)T)$.

\section{Numerical Results}
%On a plusieurs scéanar + On choisi des drones uniquement
Here, we assess the performance of our proposed MAP deployment method on dynamic scenarios at different scales.
%On défini les scenario
The first scenario, named \textbf{SmallScale}, is a small scale deployment of $10$ UEs randomly and $4$ MAPs moving through 12 positions at 3 different altitudes (i.e. $15$, $35$ and $50\m$) in a $100\m$ by $100\m$ area. In this scenario, we can easily compare our method with brute force mechanism, named \textbf{Exhaustive}, without too high computational time. The \textbf{MediumScale} scenario is made with $40$ UEs and $10$ MAPs moving through $27$ positions at $3$ altitudes in a $200\m$ by $200\m$ area. The UEs are deployed with uniform probability and move randomly over the cell. %The UEs chose randomly a random traffic following a gaussian distribution bit/s each iteration and then simulate dynamic rate demand.
The UEs' traffic follows a Poisson distribution bit/s and simulates dynamic rate demand.
%On compare avec benchmark full random pour chaque scénar
Both scenarios include a baseline random algorithm, named \textbf{Random}, where the deployment decision and the chosen location follow a uniform law $U(0,1)$. %The exhaustive benchmark is only computed for the small scale scenario. 
%For that benchmark, for each iteration, each UAV chooses a random position and we observe the network performance. When a better combination is found to the problem \ref{eq:P1}, the solution is updated.\\
Thus each UAV chooses a random location and when a better combination is found, the solution is updated.

%On défini les specs pour les drones maintenant
Concerning MAPs, each UAV has a coverage range defined by the aperture angle of its antenna to ground UEs and its altitude. At most $K=10$ UAVs can de deployed and each UAV can connect to at most $N_{i}=10$ UEs.
%Each UAV can connect with $N_{i}$=10 UEs with $K=10$ available UAVs.
%To model UAV energy consumption, we use the empirical model studied in \cite{Abeywickrama2018}.
%by Abeywickrama et Al.% .
% Let $e_{i}$ be the energy consumed by MAP $i$ to move from position $k$ to $p$:
% \begin{equation}\label{kei}
%     e_{i} = E_{at} + E_{Ho} + E_{Ht} + E_{Vt} + E_{t_{o}},
% \end{equation}
%\footnote{$E_{at}= 29.027t_{at}-0.087$,$E_{H_{o}} = (4.917H_{o}+275.204)t_{H_{o}}$,$E_{Ht} = 308.709t_{h}-0.852$,$E_{Vt} = 315h-211.261, h\geq 0$,$E_{Vt} = 68.956h-65.183, h\leq 0$,$E_{to} = -0.432V_{to}^2+3.786V_{to}-1.224$
%where $t_{at}$ is the time spent with pale working, $H_{o}$ the hovering height, $t_{H_{o}}$ the hovering time, $t_{h}$ the horizontal flight time, h the height difference between $k$ and $p$ and $V_{to}$ the take off speed}. 
We fix the rent cost $c_{i, 0}=c_0=1,~\forall i$, so that we can omit it from the optimization in Eq. \eqref{costi}. Table \ref{tab:channel} gives an overview of the simulation parameters.

\begin{table}[!t]\label{table12}
\caption{Simulation Parameters}
    \centering
    \scalebox{\tablescale}{
    \renewcommand{\arraystretch}{1.1}
    \begin{tabular}{l||c|c }
    \hline
    %\multicolumn{1}{|c|} {\textbf{Scenario Parameters}} & \multicolumn{1}{c|} {\textbf{Small Scale}} & \multicolumn{1}{c|} {\textbf{Medium Scale}} \\
    \textbf{Scenario Parameters} & \textbf{Small Scale} & \textbf{Medium Scale}\\
    \hline
     Cell size & $100 \times 100\m$ &$ 200 \times 200\m$ \\
      %\hline
      %UE Deployment & \multicolumn{2}{c}{Uniform Probability} \\
      \hline
      Number of UEs & $10$ & $40$ \\
      \hline
      Number of positions & $4 \times 3 = 12$ & $9 \times 3 = 27$ \\
      \hline
      UE Mobility & \multicolumn{2}{c}{Random Walk}\\
      \hline
      Avg. traffic demand $D_j(t)$ & \multicolumn{2}{c}{$200\Mbps$} \\
    \hline
    \hline
    \textbf{Channel Parameters} & \textbf{MBS} & \textbf{BS/UAV} \\
      \hline
      Carrier Frequency $f_{c}$ & $2\GHz$ & $28\GHz$\\
      \hline
      Bandwidth & $10\MHz$ & $500\MHz$ \\
      \hline
      Thermal Noise $N_{0}$ & \multicolumn{2}{c}{$-174\dBm/\Hz$} \\
      \hline
      Shadowing power $\sigma^2$ & $9\dB$ & $12\dB$\\
      \hline
      Transmit Power & $46\dBm$ & $20\dBm$ \\
      \hline
      Antenna Gain & 17 dBi & Directive \cite{Sana2021} \\
      \hline
      UAV Aperture & \multicolumn{2}{c}{$120\deg$}\\
      \hline
      Altitude & \multicolumn{2}{c}{$[10,35,50] \m$}\\
      \hline
    \end{tabular}}
    \label{tab:channel}
\end{table}

\subsection{Drone Deployment}
\begin{figure}[!htp]
%\vspace{-0.3cm}
\centering
\includegraphics[width=\columnwidth]{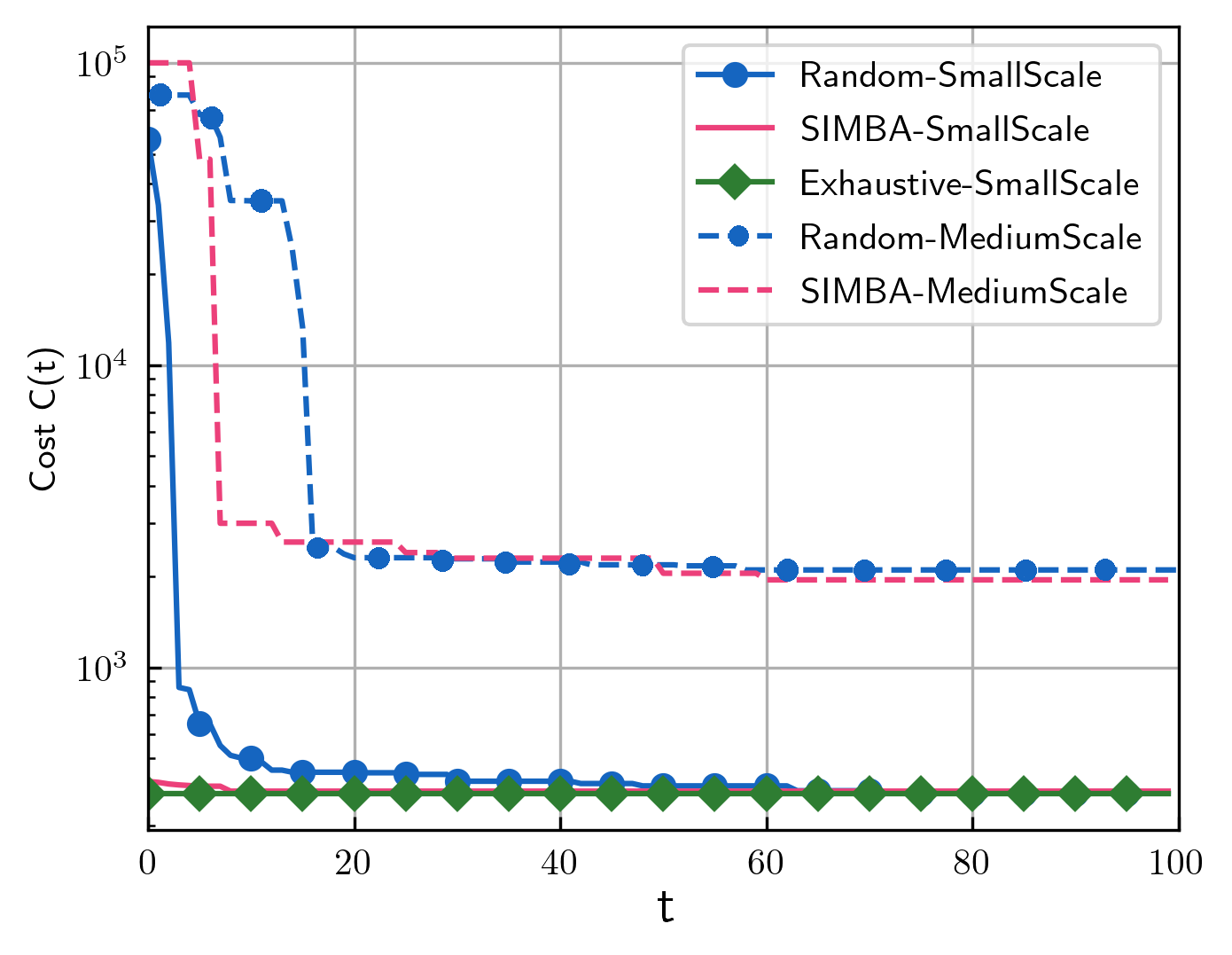}
\vspace{-0.6cm}
\caption{MAP deployment cost for each method and for both scenarios as a function of the number of iterations.} 
%\textcolor{red}{l'iteration a un nom dans le papier qui est $T$ et cost aussi qui est $C(t)$ met les dans xlabel et ylabel e.g. cost C(t). Tout ca evite la confusion, par ex iteration c'est T ou Monte-Carlo. Aussi si possible met des markeurs sur les courbes ça permet de s'en sortir quand tu imprimes en noir et blanc!}}
\label{fig:cost_graph}
\end{figure}

%Contexte de la simulation
We set $M=10$ and $T=100$ in \texttt{SIMBA} and average the results over $30$ Monte-Carlo simulations. 
%Given the probabilistic aspect of our solution, we averaged the results over 30 simulations of 100 iterations each. 
We conveniently fix the QoS's target of \eqref{eq:C5} to $Q_j=100\%$ for \textbf{SmallScale} and $Q_j = 85\%$ for \textbf{MediumScale} due to limited radio resources. 
%meaning that a deployment is considered a solution when $100\%$ and $85\%$ of the UEs are well deserved (i.e. their data rate is higher than the required data rate).
%On résout le problème => Cost minimization Figure 1
%On converge vite => Flexibility
We first assess the deployment cost of our proposed solution compared to the two benchmarks as shown in Fig. \ref{fig:cost_graph}. Our algorithm converges faster than a naive random approach to a close optimal-solution. In a high mobility context, it is important to obtain a flexible algorithm that finds a solution faster than the network changes.%that has time to find a solution before the network changes too much.

\begin{figure}[!htp]
\centering
\includegraphics[width=\columnwidth]{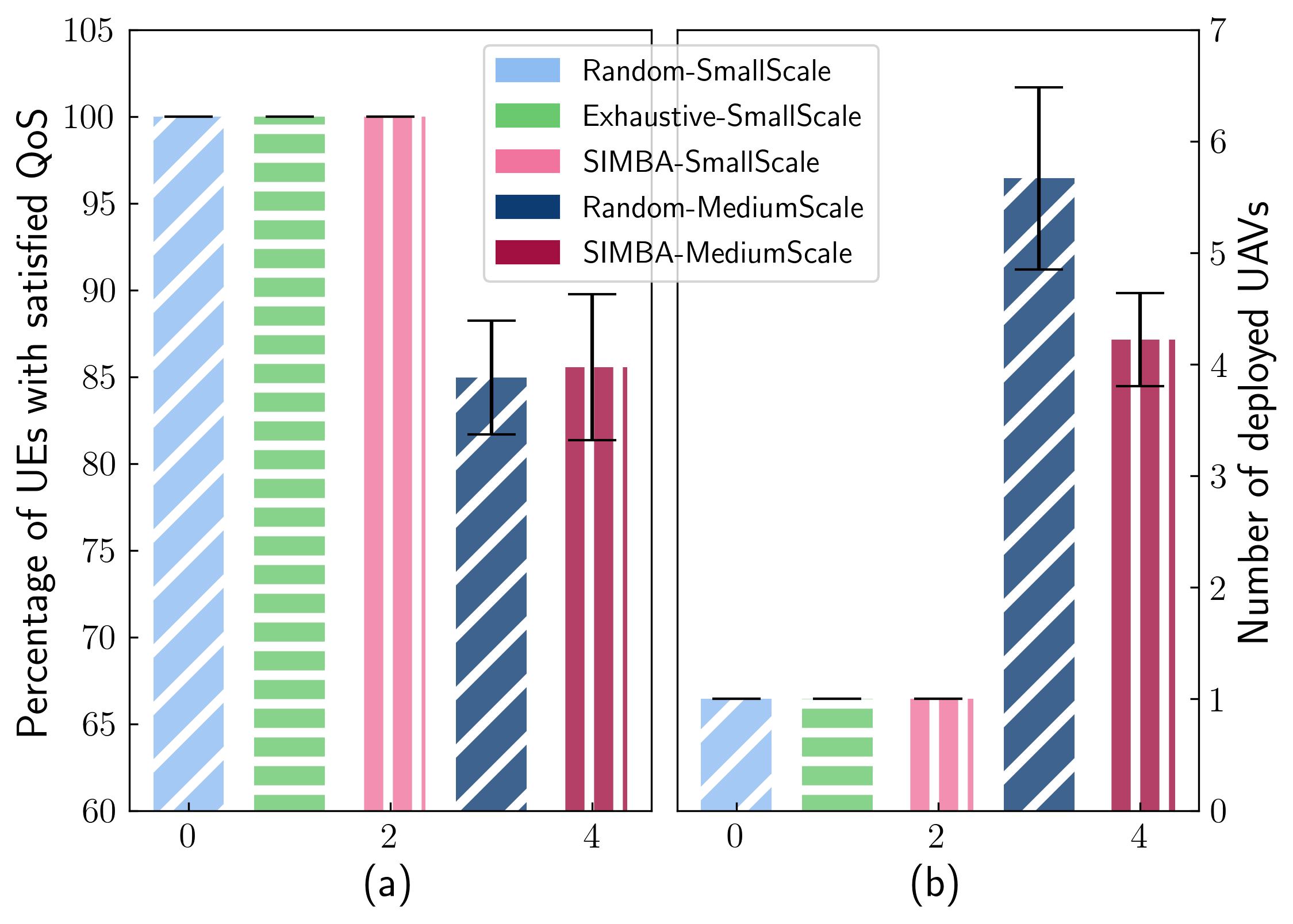}
\vspace{-0.6cm}
%\caption{Number of deserved UE, QoS and number of UAV deployed for both scenario and both method}
\caption{(a) Average percentage of UEs with satisfied QoS and (b) average number of UAVs deployed for both scenarios and for each method.} %\textcolor{red}{Cette figure n'est pas claire, on voit pas toute suite que c'est 2 figures différentes colée, on ne sait pas à quelle bar s'applique les labels à gauche et droite. c'est pas intuitif. fait 2 schéma bien séparée quitte à diminuer la largeur des bar comme dans l'ex}}
\label{fig:perf_graph3.1}
\vspace{-0.3cm}
\end{figure}

%On suit la user demand => On a un level de QoS au dela d'un nombre de bien connecté
%Important car la demande évolue dans le réseau
Moreover, our algorithm guarantees not to deploy more drones than needed, which may imply a high cost. Meanwhile, as shown in Fig. \ref{fig:perf_graph3.1}.a, our solution ensures and guarantees the targeted QoS for UEs in both small and medium scale scenarios with less MAPs compared to a naive approach as illustrate \ref{fig:perf_graph3.1}.b. %It also shows the importance of taking interference into account in our network. %\textcolor{red}{With a small group of UEs (compared to an ultra-dense city) the number of well served UEs is $11\%$ less than the number of UEs connected to the network.} 
Finally, as we added more potential locations for UAVs in \textbf{MediumScale} scenario, our solution is better at identifying the best combinations than a random naive approach and faster than an exhaustive solution, thus, guaranteeing a near-optimal solution with $4.22$ UAVs deployed in average. %over 30 simulations.

%================Graph1=========
%================Graph2=========

\subsection{User Association}
\begin{figure}[!t]
%\vspace{-0.3cm}
\centering
\includegraphics[width=\columnwidth]{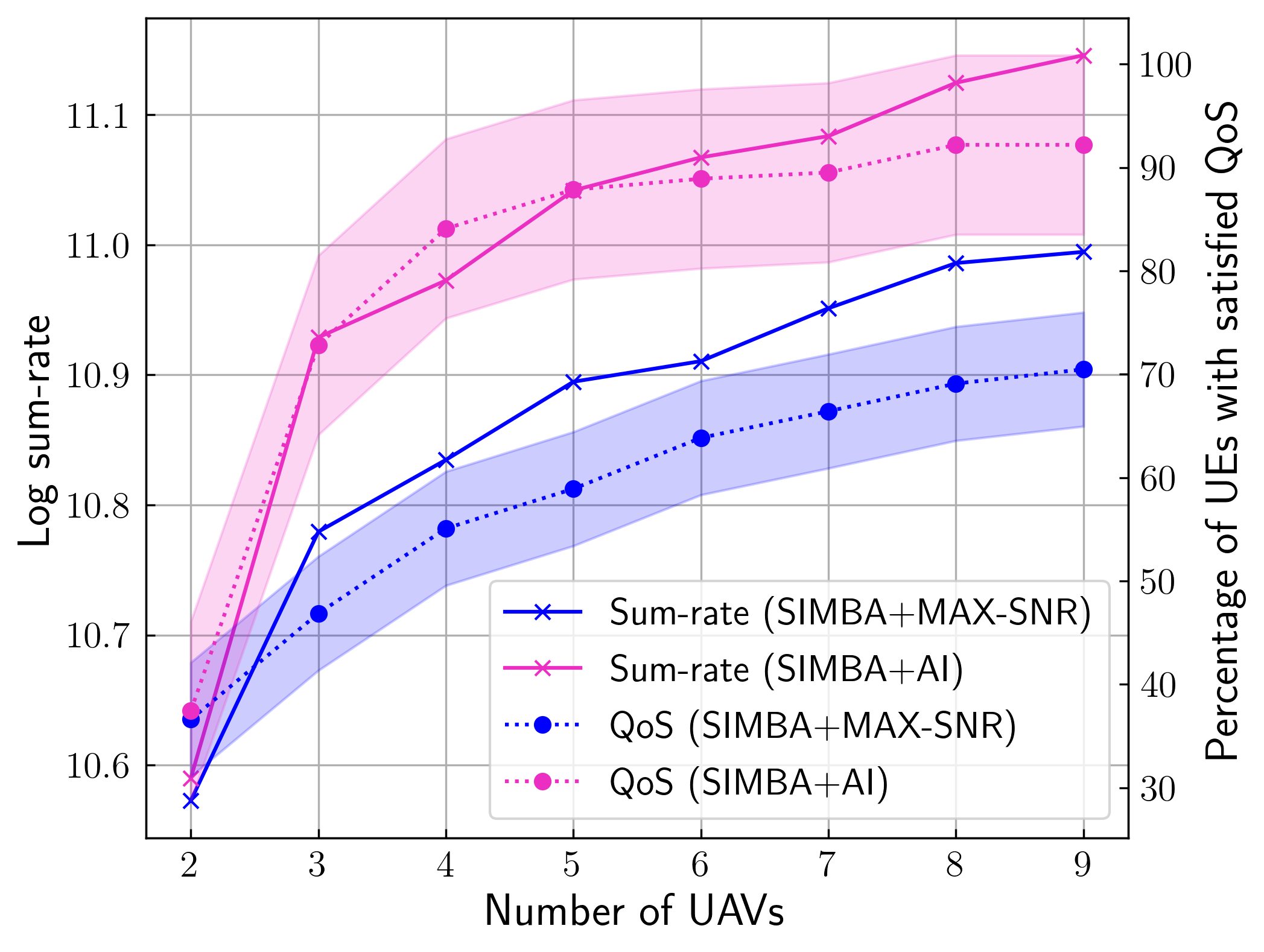}
\vspace{-0.6cm}
\caption{Average network log sum-rate and average percentage of UEs with QoS satisfaction as a function of the number of deployed UAVs for \textbf{MediumScale} scenario for both UE association algorithms.} %\textcolor{red}{log sum-rate en yaxis}}
\vspace{0.05cm}
\label{fig:graph_density_sum}
\end{figure}

%Context de la simulation
To train the user association policy, we use the MARL framework described in section \ref{user_association}. All simulation results are plotted for a learning rate $\mu=10^{-4}$ and a discount factor $\gamma=0.6$. The MARL agents are trained for $T_e=3000$ episodes. Agents are trained with $\alpha=1$ for the $\alpha$-fair utility function, meaning that the agents are trained on a fair sum-rate setting. Note that all the hyperparameters were determined empirically.
% 1. On regarde en densifiant les UAV. En densifiant le nombre de drone pour une situation donnée, on va offrir plus d'opportunité pour les utilisateurs de trouver une connection satisfaisante => Mais plus de chance d'avoir des interferences. On note un seuil de QoS ou le déploiement de drone suplémentaire introduirait un nouveau cout pour un faible gain de qos. Scénario cas classique mais on prend un scénario plus compliqué avec densification :
% 2. On densifie les users autour d'un point dans le réseau => Fléxibilité, situation plus rare et plus compliquée à gérer => Opportunité d'interfernce encore plus forte car plus d'utilisateurs pour une même antenne. => Offloading du réseau 
% 
We perform several deployments and trainings with increasing number of MAPs and compare the proposed solution to the MAX-SNR-based approach for \textbf{MediumScale} scenario. Here, results are averaged over $300$ simulations with $T=150$ iterations.
Fig. \ref{fig:graph_density_sum} shows the impact of increasing the number of UAVs on the network performance. We observe that, for a number of deployed UAVs greater than tree, our proposed approach increases the log network sum-rate by $1.5\%$, implying a network sum-rate enhancement by nearly $30\%$ compared to MAX-SNR algorithm.
%\textcolor{red}{The AI model increases the log network sum-rate by $1\%$, implying a network sum-rate by nearly $12\%$ and UEs QoS by $23\%$, highlighting the UE demand awareness.} 
Moreover, for the given scenario, Fig. \ref{fig:graph_density_sum} illustrates the trade-off between drone deployment and UE QoS. With more than $4$ UAVs deployed, the log sum-rate barely varies. This result confirms that when interference is taken into account, increasing the number of access points does not necessarily implies better UE's QoS at the risk of increasing the deployment cost. %This number of deployed UAVs confirms the effectiveness of the MAP deployment presented in Fig. \ref{fig:perf_graph3.1} with a mean of 4.22 UAVs deployed. A higher number of deployed MAPs introduces a higher cost for less improvement in UE performance.

\begin{figure}[!t]
%\vspace{-0.3cm}
\centering
\includegraphics[width=\columnwidth]{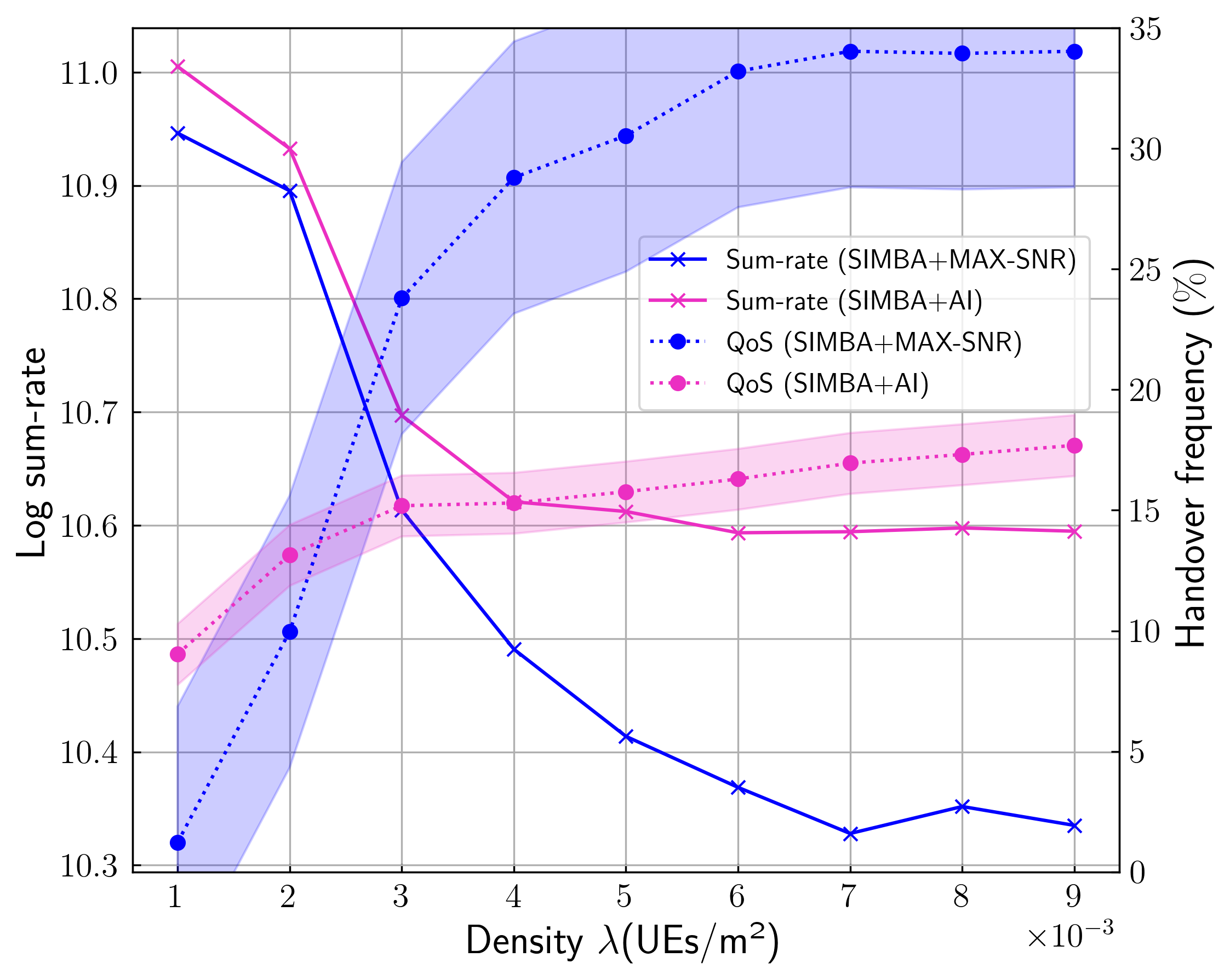}
\vspace{-0.6cm}
\caption{Average network log sum-rate and average handover frequency as a function of the user density.}
\vspace{0.05cm}
\label{fig:graph_density_all3}
\end{figure}

Next, in the \textbf{MediumScale} scenario, we increase the user density $\lambda (\mathrm{UEs}/\mathrm{m}^2)$ to show the effectiveness of our solution in this complex setting. % for dense scenarios and complex configurations. 
Fig. \ref{fig:graph_density_all3} compares the average handover frequency and the network log sum-rate as a function of user density $\lambda$. The increase of $\lambda$ ultimately increases the number of handovers frequency for the MAX-SNR algorithm as multiple UEs compete for the same resources. In contrast, our proposed solution guarantees stable performance due to its capability to to balance the network load, especially in dense deployment scenarios. %, which shows the effectiveness of MAPs for the dense deployment scenario. 
As shown in Fig. \ref{fig:graph_density_all3}, our proposed solution improves the log network sum-rate by $4\%$, which implies an increase in network sum-rate by $60\%$ compared to a MAX-SNR algorithm, in particular for dense deployment scenario (e.g. $\lambda=9\times10^{-3} \mathrm{UEs}/\m^2$).

%To conclude, our proposed solution handles a standard uniform network configuration, often studied in the state of the art, with low-cost MAP deployment and user association, with a $30\%$ increase in sum-rate and a $22\%$ improvement in UE QoS. In addition, tested on highly dynamic scenarios with dense deployment and mobility, our solution provides up to increases $60\%$ network sum-rate increase compared to MAX-SNR-based approach.

%================Graph4=========
%================Graph3=========
%================Graph2=========
%\begin{figure}[!htp]
%\vspace{-0.3cm}
%\centering
%\includegraphics[width=\columnwidth]{./Images/graph_density_all2.png}
%\vspace{-0.3cm}
%\caption{}
%\vspace{0.05cm}
%\label{fig:graph_density_all}
%\end{figure}
%================Graph5=========
%\begin{figure}[!htp]
%\vspace{-0.3cm}
%\centering
%\includegraphics[width=\columnwidth]{./Images/graph_density_hand3.png}
%\vspace{-0.3cm}
%\caption{Number of handovers for the whole simulation versus the number of deployed UAV for the second scenario for AI and SNR user association algorithms}
%\vspace{0.05cm}
%\label{fig:graph_density_hand}
%\end{figure}
%==========================================================

\section{Conclusion}
In this paper, we proposed an algorithm to solve the joint problem of MAP deployment and user association while taking into account UE mobility, UE demand and network interference. The algorithm finds the MAP positions in a 3D space and optimizes the user association for a given network configuration. The proposed algorithm is the first step for more complex scenarios. We will consider path planning optimization to exploit MAP connectivity while moving within the network, include the backhaul constraint and optimization and extend our study to more realistic configurations with aerial and ground base stations.

\section*{Acknowledgment}
This work was supported by the European Union H2020 Project \mbox{DEDICAT 6G} under grant no. 101016499. The contents of this publication are the sole responsibility of the authors and do not in any way reflect the views of the EU.

\bibliographystyle{ieeetr}
\bibliography{./Biblio}
\end{document}